\documentclass[fleqn,usenatbib]{mnras}

\usepackage{newtxtext,newtxmath}

\usepackage[T1]{fontenc}

\DeclareRobustCommand{\VAN}[3]{#2}
\let\VANthebibliography\thebibliography
\def\thebibliography{\DeclareRobustCommand{\VAN}[3]{##3}\VANthebibliography}

\usepackage{graphicx}	
\usepackage{amsmath}	

\newcommand{\galone}{Mrk 1434}
\newcommand{\galtwo}{SDSS J1213}
\newcommand{\galthree}{SDSS J1221}

\newcommand{\mrks}{Mrk 1434 X-S}
\newcommand{\mrkn}{Mrk 1434 X-N}

\title[X-ray Sources in Dwarf Galaxies]{Multiwavelength Scrutiny of X-ray Sources in Dwarf Galaxies: ULXs versus AGN}

\author[E. Thygesen et al.]{Erica Thygesen,$^{1}$\thanks{E-mail: thygesen@msu.edu}
Richard M. Plotkin,$^{2,3}$\thanks{E-mail: rplotkin@unr.edu}
Roberto Soria,$^{4,5,6}$
Amy E. Reines,$^{7}$
Jenny E. Greene,$^{8}$ \newauthor
Gemma E. Anderson,$^{9}$
Vivienne F. Baldassare,$^{10}$
Milo G. Owens,$^{2}$
Ryan T. Urquhart,$^{1}$
Elena Gallo,$^{11}$ \newauthor
James C. A. Miller-Jones,$^{9}$
Jeremiah D. Paul,$^{2}$
and Alexandar P. Rollings$^{2}$
\\
$^{1}$Center for Data Intensive and Time Domain Astronomy, Department of Physics and Astronomy, Michigan State University, East Lansing, MI 48824, USA\\
$^{2}$Department of Physics, University of Nevada, Reno, NV 89557, USA\\
$^{3}$Nevada Center for Astrophysics, University of Nevada, Las Vegas, NV 89154, USA\\
$^{4}$College of Astronomy and Space Sciences, University of the Chinese Academy of Sciences, Beijing 100049, China\\
$^{5}$INAF - Osservatorio Astrofisico di Torino, Strada Osservatorio 20, 10025, Pino Torinese, Italy\\
$^{6}$Sydney Institute for Astronomy, School of Physics A28, The University of Sydney, Sydney, NSW 2006, Australia\\
$^{7}$eXtreme Gravity Institute, Montana State University, Bozeman, MT, USA\\
$^{8}$Department of Astrophysical Sciences, Princeton University, Princeton, NJ 08544, USA\\
$^{9}$International Centre for Radio Astronomy Research, Curtin University, GPO Box U1987, Perth, WA 6845, Australia\\
$^{10}$Department of Physics and Astronomy, Washington State University, Pullman, WA 99163, USA\\
$^{11}$Department of Astronomy, University of Michigan, 1085 S University, Ann Arbor, MI 48109, USA\\
}

\date{Accepted XXX. Received YYY; in original form ZZZ}

\pubyear{2023}

\begin{document}
\label{firstpage}
\pagerange{\pageref{firstpage}--\pageref{lastpage}}
\maketitle

\begin{abstract}
Owing to their quiet evolutionary histories, nearby dwarf galaxies (stellar masses $M_\star \lesssim 3 \times 10^9 M_\odot$) have the potential to teach us about the mechanism(s) that ‘seeded’ 
the growth of supermassive black holes, and also how the first stellar mass black holes formed and interacted with their environments.
Here, we present high spatial-resolution observations of three dwarf galaxies in the X-ray (Chandra), the optical/near-infrared
(Hubble Space Telescope), and the radio (Karl G.\ Jansky Very Large Array).  These three galaxies were  previously identified as hosting candidate active galactic nuclei on the basis of lower resolution X-ray imaging. 
With our new observations, we find that X-ray sources
in two galaxies (SDSS J121326.01+543631.6 and SDSS J122111.29+173819.1) are off nuclear and lack corresponding radio
emission, implying they are likely luminous X-ray binaries. The third galaxy (Mrk 1434) contains two X-ray sources (each with $L_{\rm X} \approx 10^{40}$ erg s$^{-1}$) separated by 2\farcs8, has a low-metallicity (12 + log (O/H) = 7.8), and emits nebular \ion{He}{II} $\lambda$4686 line emission. The northern
source has spatially coincident point-like radio emission at 9.0 GHz and extended radio emission at 5.5 GHz. We discuss
X-ray binary interpretations (where an ultraluminous X-ray source blows a `radio bubble') and active galactic nucleus
interpretations (where a $\approx 4\times10^5 M_\odot$ black hole launches a jet). In either case, we find that the \ion{He}{II} emission cannot
be photoionised by the X-ray source, unless the source was $\approx$30–-90 times more luminous several hundred years ago.

\end{abstract}




\section{Introduction}
\label{sec:intro}

There is abundant evidence that supermassive black holes (SMBHs; $10^6 \lesssim M_{\rm BH} \lesssim 10^9~M_\odot$) ubiquitously exist at the centres of large galaxies \citep[e.g.,][]{kormendy13}, some of which accrete and shine as active galactic nuclei (AGNs).  Some lower-mass dwarf galaxies (which we define by stellar masses $M_\star \lesssim 3 \times 10^9 M_\odot$) are known to host nuclear black holes (e.g., \citealt{filippenko03, barth04, reines11, reines13, schramm13, moran14, sartori15, mezcua16, mezcua18, pardo16, ho16, chen17, chilingarian18, nguyen19, baldassare20, martinez-palomera20, cann21, schutte22}), with some mass estimates as low as $M_{\rm BH} \approx$10$^4$ $M_\odot$  \citep[e.g.,][]{baldassare15, woo19}.   These black holes represent the lower-mass end of the SMBH population, and we  refer to them here as `massive black holes' (mBHs; $10^4 \lesssim M_{\rm BH} \lesssim 10^6~M_\odot$). An actively accreting mBH can  affect how dwarf galaxies provide feedback to their larger scale environments \citep[e.g.,][]{dashyan18, trebitsch18, mezcua19}, and more generally,  mBHs  represent a phase that nuclear black holes must pass through as they grow to SMBH sizes over cosmological time scales \citep[e.g.,][]{volonteri10}. Given that dwarf galaxies have had relatively quiet evolutionary histories, constraining the fraction of dwarf galaxies hosting mBHs in the local Universe, along with the mBH mass distribution, may provide constraints on the mechanism(s) that formed the first black holes in the Universe \citep[e.g.,][]{ricarte18, inayoshi20, volonteri21}.  The fraction of dwarf galaxies hosting an mBH is still relatively unknown, with current empirical constraints implying  $\gtrsim30-50$\% occupation \citep{miller15, gallo19, greene20}.  

Stellar mass black holes ($M_{\rm BH} \approx$10 $M_\odot$) and neutron stars are also observed within some dwarf galaxies in the form of X-ray binaries (XRBs).   XRBs serve as probes of stellar populations within galaxies, with the number and/or luminosity of XRBs expected to scale with the star formation rate, stellar mass, and metallicity of the host galaxy \citep[e.g.,][]{grimm03, gilfanov04, linden10, mineo14, lehmer21}.  Intriguingly, lower-metallicity galaxies appear to contain an excess of luminous XRBs compared to Solar-metallicity galaxies \citep{prestwich13, brorby14, douna15, ponnada20, lehmer21}, which may be a consequence of lower-metallicity progenitor stars having weaker stellar winds, and therefore producing black hole remnants that are more numerous and/or more massive (e.g., \citealt{heger03, mapelli10}). Besides tracing stellar populations, the energy output from XRBs can also provide feedback to their host galaxies.   For example, line emission from the high-ionisation \ion{He}{ii} $\lambda$4686  line ($\chi_{\rm ion} = 54.4$ eV) has been observed from some ultraluminous X-ray sources (ULXs),\footnote{We define ULXs as having X-ray luminosities $L_X > 10^{39}$ erg s$^{-1}$.  ULXs  are most commonly interpreted as super-Eddington neutron star or black hole XRBs \citep[see, e.g.,][]{feng11, kaaret17}.} 
which is often interpreted as an X-ray photoionised nebula \citep{pakull86, moon11}.   Extrapolating such ULX phenomenology in the local Universe to higher redshifts, XRBs could have contributed to the X-ray heating of the intergalactic medium during the Epoch of Reionisation and Cosmic Dawn \citep[e.g.,][]{mirabel11, ponnada20}. 
Thus, characterising both the XRB and mBH populations in nearby dwarf galaxies, particularly as a function of host galaxy metallicity, is important for understanding the formation of the first black holes and galaxies in the Universe.

X-ray observations are commonly used to  identify accreting compact objects, since hard X-ray emission ($\gtrsim$1--2 keV) is a universal signature of accretion. However, in several cases, it is very challenging to determine the mass of an accreting object via X-ray observations alone.  In particular, both a rapidly-accreting XRB and a weakly-accreting mBH/SMBH  can have comparable X-ray luminosities in the $10^{39}-10^{41}$ erg s$^{-1}$ range, and they can also display similar X-ray spectral shapes (below $\approx$50 keV).  Folding in multiwavelength information is therefore essential for differentiating between rapidly accreting XRBs and weakly-accreting mBHs/SMBHs.  It is well established that weakly accreting SMBHs (i.e., low-luminosity AGNs with $L_{\rm bol} \lesssim 0.01 L_{\rm  Edd}$, where $L_{\rm bol}$ is the bolometric luminosity and $L_{\rm Edd}=1.3\times10^{38} \left[M_{\rm BH}/M_\odot\right]$ erg s$^{-1}$ is the Eddington luminosity)  emit compact, usually unresolved, radio emission with a flat spectrum \citep[$f_\nu \propto \nu^{-\alpha}$, where $f_\nu$ is the radio flux density at frequency $\nu$, and the radio spectral index $\alpha=0$ for a flat spectrum;][]{ho08}.  Such unresolved, flat spectrum radio emission is usually interpreted as a partially self-absorbed synchrotron jet \citep{blandford79}.  Meanwhile, rapidly accreting XRBs do not  launch  jets that would be detectable beyond distances of a few Mpc \citep{fender04}.  Thus, the presence of unresolved radio emission has the potential to exclude hard X-ray sources as rapidly accreting XRBs. 

In this paper, we present high spatial-resolution X-ray (\textit{Chandra}), optical/near-infrared (\textit{Hubble Space Telescope; HST}), and radio observations (Karl G.\ Jansky Very Large Array; VLA) of three nearby dwarf galaxies that each host at least one hard X-ray point source.  These three targets  were initially identified as AGN candidates by \citet{lemons15}, but with the caveat that the positions of their X-ray sources were poorly determined.  From the multiwavelength data presented here, we better locate the positions of the X-ray sources within these three galaxies, and we attempt to constrain the nature of each source (i.e., XRB or mBH).  In Section \ref{sec:obs} we detail our sample selection and data reduction. We outline our results in Section \ref{sec:results}, followed by a discussion in  Section \ref{sec:disc}. Our conclusions are  presented in Section \ref{sec:conc}.  Unless stated otherwise,  uncertainties are reported at the 68\% confidence level.

\section{Observations and Data Reduction}
\label{sec:obs}

\subsection{Sample}
\label{sec:obs:sample}

Our three targets were selected from the survey by \cite{lemons15}, who cross matched  $\sim$44,000 nearby dwarf galaxies ($z<0.055$) from the NASA-Sloan Atlas\footnote{\url{http://www.nsatlas.org/}} to the \textit{Chandra} Source Catalog (CSC Release 1.1; \citealt{evans10}).  They identified 19 galaxies with hard X-ray point sources (2--7 keV), of which 10 contained an X-ray source positionally consistent with the galaxy optical centre (given positional uncertainties, we note that not every galaxy has a well defined nucleus).  They presented these 10 galaxies as AGN candidates.\footnote{Since publication of \citet{lemons15}, there is new theoretical evidence that  mBHs do not need to reside in the nucleus \citep[e.g.,][]{bellovary19}.}  

\begin{table}
    \centering
    \caption{Properties of the three dwarf galaxies in our sample. Column 1:  galaxy names.  The full designations of the second and third galaxies are SDSS J121326.01+543631.6 and J122111.29+173819.1. Column 2: distances to each galaxy, assuming $H_0=73$ km s$^{-1}$ Mpc$^{-1}$ for \galone\ and \galtwo, and using the Tully-Fisher relation for \galthree  \citep{kashibadze20}. Column 3: stellar masses, following the methodology of \citet{reines15}. Column 4: logarithm of star formation rates, based on far-ultraviolet and infrared luminosities \citep{hao11, kennicutt12}.  Column 5: metallicities when available in the literature (taken from \citealt{shirazi12} for \galone\ and \citealt{zhao13} for \galthree).}
    \label{tab:properties}
    \renewcommand{\arraystretch}{1.} 
    \begin{tabular}{ccccc}
        \hline
        Name & D     & $\log M_\star$ &  $\log$ SFR          & 12+$\log\left(O/H\right)$ \\
             & (Mpc) & ($M_\odot$)   &  ($M_\odot$ yr$^{-1}$) &    \\ 
        (1)  &  (2)  & (3)           &  (4)                  &  (5)   \\
         \hline
        \galone    &  30.7  &  6.6 & $-$0.9 & 7.8 \\
        \galtwo     &  32.7  & 7.3  & $-$2.2  & \ldots \\
        \galthree   & 16.1  &  8.0& $-$1.5 & 8.3 \\
         \hline
    \end{tabular}
\end{table}

\textit{Chandra} provides exquisite spatial resolution ($\approx$0\farcs4) for targets located at the telescope's aimpoint, but the resolution degrades for sources located farther away.   Of the 10 AGN candidates in \citet{lemons15}, they found that four galaxies contain X-ray sources that  are far enough  from the aimpoint to have large positional uncertainties ($>$5$\arcsec$, which is comparable to the projected size of the entire dwarf galaxy).  Of these four galaxies, three contained X-ray sources with hard  X-ray luminosities $>$3$\sigma$ ($>$1.2 dex) larger than expected from the galaxy-wide contribution from X-ray binaries, given the stellar mass and star formation rate of each galaxy (see Sections 4.3 and 4.4 of \citealt{lemons15}).  These three galaxies include:  \galone\ ($z=0.00747$), SDSS J121326.01+543631.6 ($z=0.00797$; hereafter \galtwo), and SDSS J122111.29+173819.1 ($z=0.00699$; hereafter \galthree; see Table~\ref{tab:properties}).  Of particular interest, \galone\ is a metal-poor galaxy ($12+\log\left(O/H\right)=7.8$; \citealt{shirazi12}) and its optical spectrum from the Sloan Digital Sky Survey (SDSS; \citealt{york00}) shows nebular \ion{He}{ii} line emission \citep{shirazi12}.

To better constrain the locations of the X-ray sources relative to their host galaxies, we obtained new \textit{Chandra} X-ray and \textit{HST} optical/near-infrared observations for these three galaxies.  We also obtained new VLA radio observations for one target,  \galtwo, while archival VLA data were already available for the other two sources.   We adopt distances for each galaxy based on their redshifts, using $H_0=73$ km s$^{-1}$ Mpc$^{-1}$, except for \galthree, which is a member of the Virgo cluster (VCC 459).  For this galaxy, we use a distance of 16.1 Mpc based on the Tully-Fisher relation \citep{kashibadze20}.  For all three galaxies, we adopt star formation rates from \citet{lemons15}, which are based on (dust-corrected) far-ultraviolet and infrared luminosities and the relationships from \citet{hao11} and  \citet{kennicutt12}.  For \galthree, we scale the star formation rate from \citet{lemons15} to 16.1 Mpc.  For stellar mass estimates, following \citet{reines15}, we use the colour-dependent mass-to-light ratios from \citet{zibetti09}.

\subsection{\textit{Chandra}} 
\label{sec:obs:chandra}

We obtained new \textit{Chandra} observations 
(Cycle 17; PI Plotkin) with each galaxy centred at the aimpoint of the S3 chip of the Advanced CCD Imaging Spectrometer \citep[ACIS;][]{garmire03}. Data were telemetered in {\sc VFAINT} mode.  
\textit{Chandra} data reduction was carried out using the Chandra Interactive Analysis of Observations ({\sc ciao}) software version 4.13 \citep{fruscione06} and {\tt caldb} v4.9.5. The \textit{Chandra} data were reprocessed using {\tt chandra$\_$repro} to create new level 2 event files and bad pixel files, and to apply the latest calibration files.  We then searched for background flares using the {\tt deflare} script, and we did not find any periods with elevated background levels.   

\begin{table*}
    \centering
    \caption{Details of \textit{Chandra} observations.  Column 1:  name of X-ray source.  %
    Column 2: \textit{Chandra} obsID. %
    Column 3: date of observation. %
    Column 4: exposure time. %
    Columns 5 \& 6: right ascension and declination of each X-ray    source. %
    Column 7: radius of the 95\% positional uncertainty of each \textit{Chandra} source, based on Equation 5 of \citet{hong05}. %
    Column 8:  aperture corrected net count rate (in counts per ks) in the broad X-ray band (0.5-7.0 keV).  Aperture corrections of 0.90, 0.95, and 0.96 were used for \galone, \galtwo, and \galthree, respectively.  %
    Column 9: aperture corrected net count rate (in counts per ks) in the hard band (2.0-7.0 keV). Aperture corrections of 0.87, 0.93, and 0.93 were used for \galone, \galtwo, and \galthree, respectively.}
    \label{tab:chandra}
    \renewcommand{\arraystretch}{1.2} 
    \begin{tabular}{ccccccccc}
        \hline
        Source & obsID  & Date  & Exp Time & Right Ascension & Declination & $p_{\rm err}$ & Net Rate (0.5-7.0 keV) & Net Rate (2.0-7.0 keV) \\ 
               &        &       & (ks)     & (J2000)         & (J2000)     & ($\arcsec$)& (ks$^{-1}$)           & (ks$^{-1}$)  \\
        (1)    &  (2)   &   (3) &  (4)     & (5)             & (6)         &        (7)    & (8)     & (9)      \\
         \hline
    \mrkn & 18059 & 2016 Jan 26 &  5.0 & 10:34:10.19 & +58:03:49.0 & 0.35  & $8.00\pm2.22$ & $2.98\pm1.40$ \\ 
    \mrks & 18059 & 2016 Jan 26 & 5.0 & 10:34:10.11 & +58:03:46.3 & 0.36& $6.83\pm2.04$ & $2.06^{+1.36}_{-0.94}$ \\
    \galtwo & 18060 & 2016 Aug 04  & 7.0  & 12:13:26.12  &  +54:36:34.1 &  0.38 &  $2.78\pm1.10$ & $1.15^{+0.89}_{-0.60}$ \\ 
    \galthree & 18061 & 2016 Feb 13 & 7.0  & 12:21:11.00 & +17:38:18.0 &  0.33 &  $10.82\pm2.11$ & $3.46\pm1.24$ \\ 
         \hline
    \end{tabular}
\end{table*}

Next we aligned the event file astrometry to the SDSS reference frame. We first excluded  areas on each X-ray image occupied by the dwarf galaxy, so that our astrometric corrections would not be influenced by sources within each target galaxy.    We then filtered each \textit{Chandra} image to 0.5-7.0 keV and ran {\tt wavdetect} to identify X-ray point sources, adopting wavelet scales of 1,2,4,8, and 16,  setting {\tt sigthresh} to 10$^{-6}$ (i.e., approximately one false positive per chip), and using a point spread function map  (at 2.3 keV) with an enclosed count fraction (ecf) of 0.9.  The relatively large ecf was chosen to help filter out weak X-ray sources, which would not have sufficient positional accuracy for astrometric alignment.  We then cross-matched X-ray sources identified by {\tt wavdetect} to the SDSS catalog using {\tt wcs\_match}.  We found only two common X-ray/optical sources for \galone, zero common sources for \galtwo, and one common source for \galthree.  Thus, we applied a translational  astrometric correction for \galone\ ($\Delta x = 0.97, \Delta y = 1.32$ pixels) and for \galthree\ ($\Delta x = 0.01, \Delta y =0.96$ pixels) using {\tt wcs\_update}.  No astrometric correction was applied to \galtwo.  

We next re-ran {\tt wavdetect} on the aligned event files (filtered from 0.5-7 keV, now including each target dwarf galaxy) to determine positions in the aligned reference frame of X-ray sources hosted by each dwarf galaxy.  We used the same {\tt wavdetect} parameters as above, except we used ecf=0.3 when generating the point spread function map to allow the detection of fainter point sources.  {\tt wavdetect} identified two X-ray sources in \galone, one source in \galtwo, and one source in \galthree.  The positions of each X-ray source are listed in Table~\ref{tab:chandra}.  We estimated 95\%  uncertainties of each X-ray position based on the distance from the telescope aimpoint and the number of counts detected by {\tt wavdetect}, following Equation 5 in \citet{hong05}. Note, this 95\% positional uncertainty represents  the statistical error on each source.   For \galtwo\ in particular, where we could not perform an astrometric alignment of the \textit{Chandra} image, there is an additional systematic uncertainty that could be as large as 2$\arcsec$ (although 0$\farcs$8 is more typical).\footnote{\url{https://cxc.harvard.edu/cal/ASPECT/celmon/}}

We then measured the number of counts from each X-ray source using {\tt srcflux}.  We adopted circular apertures centred at each {\tt wavdetect} position  with radii of 5 pixels, except for \galone, which contains two X-ray sources, where we adopted radii of 2.5 pixels to avoid the regions from each X-ray source from overlapping. The number of background counts per pixel was estimated from nearby source-free regions of each image.   These measurements were performed in both broad (0.5-7.0 keV) and hard (2.0-7.0 keV) images, and we detected  19--73 counts from each source in the broad band and 8--23 counts in the hard band.  All X-ray detections (in all bands) are significant at the $>$99\% level according to the confidence tables in \citet{kraft91}.  

Finally, spectra were  extracted for each X-ray source using {\tt specextract} and fit using an absorbed powerlaw model ({\tt tbabs*powerlaw}) in the Interactive Spectral Interpretation System v1.6.2 ({\tt ISIS;} \citealt{houck00}), adopting Cash statistics \citep{cash79} given the relatively low number of counts per source.  We initially left the column density as a free parameter.  However, for three X-ray sources $N_H$ converged to zero, in which case we froze the value to the Galactic column density and refit the spectrum.  Model fluxes were calculated using the {\tt cflux} convolution model.   Spectral parameters and model fluxes are reported in Table~\ref{tab:chandraspec}.  

\begin{table*}
    \centering
    \caption{\textit{Chandra} spectral parameters, fluxes, and luminosities.  Column 1: name of X-ray source. %
    Column 2: column density. %
    Column 3: best-fit photon index. %
    Column 4: best-fit Cash statistic/degrees of freedom.  %
    Columns 5 \& 6: logarithms of the unabsorbed model X-ray flux and luminosity from 0.5-10 keV, estimated using the {\tt cflux} convolution model.  %
    Columns 7 \& 8: logarithms of the unabsorbed model X-ray flux and luminosity from 2-10 keV, estimated using the {\tt cflux} convolution model.}
    \label{tab:chandraspec}
    \renewcommand{\arraystretch}{1} 
    \begin{tabular}{cccccccc}
        \hline
               &                      &           &               & \multicolumn{2}{c}{Broad (0.5-10.0 keV)}  & \multicolumn{2}{c}{Hard (2.0-10.0 keV)} \\
        Source & $N_H$                & $\Gamma$  & C-stat/d.o.f. & log Flux & log Luminosity & log Flux & log Luminosity \\  
               & ($10^{20}$ cm$^{-2}$)  &           &              & (erg s$^{-1}$ cm$^{-2}$)    & (erg s$^{-1}$) & (erg s$^{-1}$ cm$^{-2}$) & (erg s$^{-1}$) \\
        (1)    &  (2)   &   (3) &  (4)     & (5)             & (6)         &        (7)    & (8)    \\
         \hline
    \mrkn & $<$56.9$^a$  & $1.3\pm0.4$    & 16.0/13 &  $-12.8\pm0.1$  & $40.2\pm0.1$   & $-13.0\pm0.2$  & $40.1\pm0.2$ \\
    \mrks & 0.6$^b$ & $1.7\pm0.4$    & 10.3/13 &  $-13.1\pm0.1$  & $40.0\pm0.1$   & $-13.3\pm0.2$  & $39.8\pm0.2$ \\
    \galtwo             & 1.4$^b$ & $1.3\pm0.5$    & 5.8/12 & $-13.3\pm0.2$  & $39.8\pm0.2$   & $-13.5\pm0.2$  & $39.6\pm0.2$ \\
    \galthree           & 2.7$^b$ & $1.6\pm0.3$    & 27.1/32  &  $-12.8\pm0.1$  & $39.6\pm0.1$   & $-13.0\pm0.1$  & $39.5\pm0.1$ \\
         \hline
    \multicolumn{8}{p{0.8\linewidth}}{$^a$Best-fit column density $N_H = 8.0 \times 10^{20}$ cm$^{-2}$, reported as an upper limit (95\% confidence level) because the uncertainty on the best-fit value extends down to the Galactic value of $0.6\times10^{20}$ cm$^{-2}$ .} \\
    \multicolumn{8}{p{0.8\linewidth}}{$^b$Column density frozen to the Galactic value during fitting, taken from \citet{dickey90}.} \\
    \end{tabular}
\end{table*}

\subsection{\textit{Hubble Space Telescope}} 
\label{sec:obs:hst}

We observed each galaxy with the Wide Field Camera 3 (WFC3) aboard \textit{HST} 
for one orbit per galaxy (PI Plotkin; program 14356).  For \galone\ and \galthree\ we observed in both the F110W and F606W filters (with the IR and UVIS channels, respectively), and for \galtwo, which is a fainter galaxy, we took observations only in the F110W filter.  Observations in each filter were taken over four dither positions, and we used the IRSUB512 subarray for \galone\ and \galthree.  Total exposure times in each filter are listed in Table~\ref{tab:hst}.
Data were downloaded from the Mikulski Archive for Space Telescopes (MAST), and individual exposures were aligned and combined using {\tt AstroDrizzle} in the {\tt DrizzlePac} software \citep{hack13}.\footnote{\url{https://hst-docs.stsci.edu/drizzpac}} 
The F110W drizzled images were created with plate scales 0\farcs06 pix$^{-1}$ for \galone\ and \galthree, and 0\farcs09 pix$^{-1}$ for \galtwo.  All F606W images have plate scales 0\farcs03 pix$^{-1}$.  

We aligned the \textit{HST} astrometry to the \textit{Gaia} Data Release 2 \citep{gaiadr2} reference frame using the {\tt tweakreg} task within {\tt AstroDrizzle} (after excluding sources falling within each galaxy).\footnote{We note that we aligned \textit{HST} images to the \textit{Gaia} frame and the \textit{Chandra} X-ray images to the SDSS frame, because we generally found a larger number of common \textit{HST/Gaia} sources vs.\ common \textit{HST}/SDSS sources (and vice-versa for \textit{Chandra}). Compared to the statistical uncertainty on each \textit{Chandra} position (0\farcs3--0\farcs4), we do not expect a meaningful offset between the absolute astrometry of SDSS vs.\ \textit{Gaia}, such that systematic uncertainties in our astrometric alignments are  dominated by the small number of sources used to apply the corrections.}  
For \galone, the  corrections resulted in astrometric shifts by  ($\Delta x = 1.8, \Delta y = 0.0$) pixels (from two common sources) and ($\Delta x = 1.9, \Delta y = 2.2$) pixels (from nine common sources) in the F110W and F606W filters, respectively.   For \galtwo, we shifted the F110W filter by  ($\Delta x = 0.6, \Delta y = 2.8$) pixels (five common sources).  Finally, for \galthree\ we could not identify enough common sources between the \textit{HST} image and the \textit{Gaia} catalog in the F110W filter (which has a smaller field of view).  So, we only aligned the F606W filter to the \textit{Gaia} frame, shifting by ($\Delta x = 0.2, \Delta y = 5.3$) pixels (four common sources), and we then aligned the F110W filter to the F606W filter (via three common sources between the two HST filters). 

\begin{table}
    \centering
    \caption{Summary of \textit{HST} observations.  Column 1: galaxy name. Column 2: date of observations. Column 3: filters used for observations.  Column 4: exposure times in the F110W/F606W filters, respectively, when both filters were used.    All observations were taken through \textit{HST} Proposal ID 14356.}
    \label{tab:hst}
    \renewcommand{\arraystretch}{1.} 
    \begin{tabular}{lccc}
        \hline
        Source & Date  & Filter & Exp. Time \\  
               &       &        & (min)     \\ 
        (1)  &  (2)  & (3)           &  (4) \\
         \hline
    \galone &  2016 Apr 16 & F110W/F606W & 8.6/30.9 \\  
    \galtwo  &  2016 Apr 16 & F110W       & 43.7 \\
    \galthree & 2016 Apr 9  & F110W/F606W & 8.6/26.9 \\
         \hline
    \end{tabular}
\end{table}

\subsection{Very Large Array} 
\label{sec:obs:vla}

\galone\ and \galthree\ both had archival datasets (PI Satyapal, 14A-358) from the VLA, while new data were obtained for \galtwo\ for this study (PI Plotkin, SH0563). All three galaxies were observed in the most extended A configuration.  Both \galone\ and \galthree\ observations were from 4.5-6.5 GHz (C band) and 8-10 GHz (X band), while \galtwo\ was observed only from 8-12 GHz.  

The Common Astronomy Software Applications ({\tt CASA}; \citealt{newcasa}) software package version 5.1 was used to carry out standard data reduction. We used 3C 286 to perform delay and bandpass calibrations, and to set the flux density scale.  Nearby phase calibrators (see Table~\ref{tab:vla}) were observed to solve for the time-dependent complex gain solutions.  Imaging was performed using the task {\tt tclean}, using two Taylor terms ({\tt nterms=2}) to account for the wide fractional bandwidth and natural weighting to maximise sensitivity. We achieved root-mean-square (rms) sensitivities ranging from 3.7 to 8.7 $\mu$Jy bm$^{-1}$ in each observing band (see Table~\ref{tab:vla}).  

\begin{table*}
    \centering
    \caption{Summary of VLA observations. %
    Column 1: galaxy name. %
    Column 2: VLA Program ID. %
    Column 3: date of observation. %
    Column 4: name of phase calibrator. %
    Column 5: central frequency of each observation. %
    Column 6: bandwidth of each observation. %
    Column 7: the time spent integrating on each galaxy.  %
    Column 8:  the size of the  (elliptical) synthesised beam along the     major and minor axes. %
    Column 9: rms noise of each image.}
    \label{tab:vla}
    \renewcommand{\arraystretch}{1} 
    \begin{tabular}{ccccccccc}
        \hline
        Source & Program    & Date  & Phase Calibrator & $\nu$ & $\Delta \nu$ & $\tau$ & $\theta_{\rm bm}$ & $\sigma_{\rm rms}$ \\  
               &            &        & (J2000)         & (GHz) & (GHz)       & (min) & ($\arcsec \times \arcsec$) & ($\mu$Jy bm$^{-1}$) \\
        (1)    &  (2)   &   (3) &  (4)     & (5)             & (6)         &        (7)    & (8)  & (9)  \\
         \hline
    \galone$^a$ & 14A-358 & 2014 Feb 24 & 1035+564   & 5.5 & 2.0 & 8.5  &     0.45$\times$0.38 & 8.7 \\
    \galone$^b$ & 14A-358 & 2014 Feb 24 & 1035+564   & 9.0 & 2.0 & 8.5 &     0.27$\times$0.24 & 8.6 \\
    \galtwo & SH0563 & 2016 Sep 30 & 1219+482    & 10.0 & 4.0 & 39.5 &     0.28$\times$0.23 & 3.7 \\
    \galthree & 14A-358 & 2014 Feb 26 & 1158+248 & 5.5 & 2.0 & 25.8 &     0.42$\times$0.38 & 5.9  \\
    \galthree & 14A-358 & 2014 Feb 26 & 1158+248 & 9.0 & 2.0 & 26.0& 0.25$\times$0.23 & 5.8 \\
    \hline
    \multicolumn{9}{p{0.75\linewidth}}{$^a$Extended radio emission detected near \mrkn\ at 5.5 GHz, with $f_{\rm int}=0.191\pm0.036$ mJy and $f_{\rm peak}=0.054\pm0.009$ mJy bm$^{-1}$.  The centroid of emission is located at RA=10$^{\rm h}$34$^{\rm m}$10.1867$^{\rm s}$ $\pm$ 0.0042$^s$, Dec=58$^\circ$03$\arcmin$49$\farcs$1481 $\pm$ 0$\farcs$0763.} \\
    \multicolumn{9}{p{0.75\linewidth}}{$^b$Point-like radio emission detected near \mrkn\ at 9.0 GHz, with $f_{\rm peak}=0.036\pm0.009$ mJy bm$^{-1}$.  The emission is located at RA=10$^{\rm h}$34$^{\rm m}$10.2045$^{\rm s}$ $\pm$ 0.0039$^{\rm s}$, Dec=58$^\circ$03$\arcmin$49$\farcs$2883 $\pm$ 0$\farcs$0460.} \\
    \end{tabular}
\end{table*}

 The only X-ray source for which we found coincident radio emission is \mrkn, where we found radio detections at both 5.5 and 9.0 GHz within the X-ray error circle.  We used {\tt imfit} to fit two-dimensional Gaussians in the image plane (at each frequency) to calculate the size of the radio structure, and to measure  peak and integrated flux densities. As discussed further in Section \ref{sec:res:mrk1434}, the 5.5 GHz emission is slightly extended (with integrated flux density $f_{\rm int} = 0.191\pm0.036$ mJy) while the 9.0 GHz is point-like ($f_{\rm peak} = 0.036\pm0.009$ mJy). The centroids of the radio emission at each frequency are offset by $0\farcs20\pm0\farcs07$.  For the other two galaxies, we place 3$\sigma_{\rm rms}$ limits on their radio flux densities.  We note that  \galthree\  displays radio emission aligned with a likely \ion{H}{ii} region toward the eastern outskirts of the galaxy that is not associated with X-ray emission, so we do not discuss that radio emission in this paper.

\section{Results}
\label{sec:results}

\begin{figure*}
    \centering
    \includegraphics[]{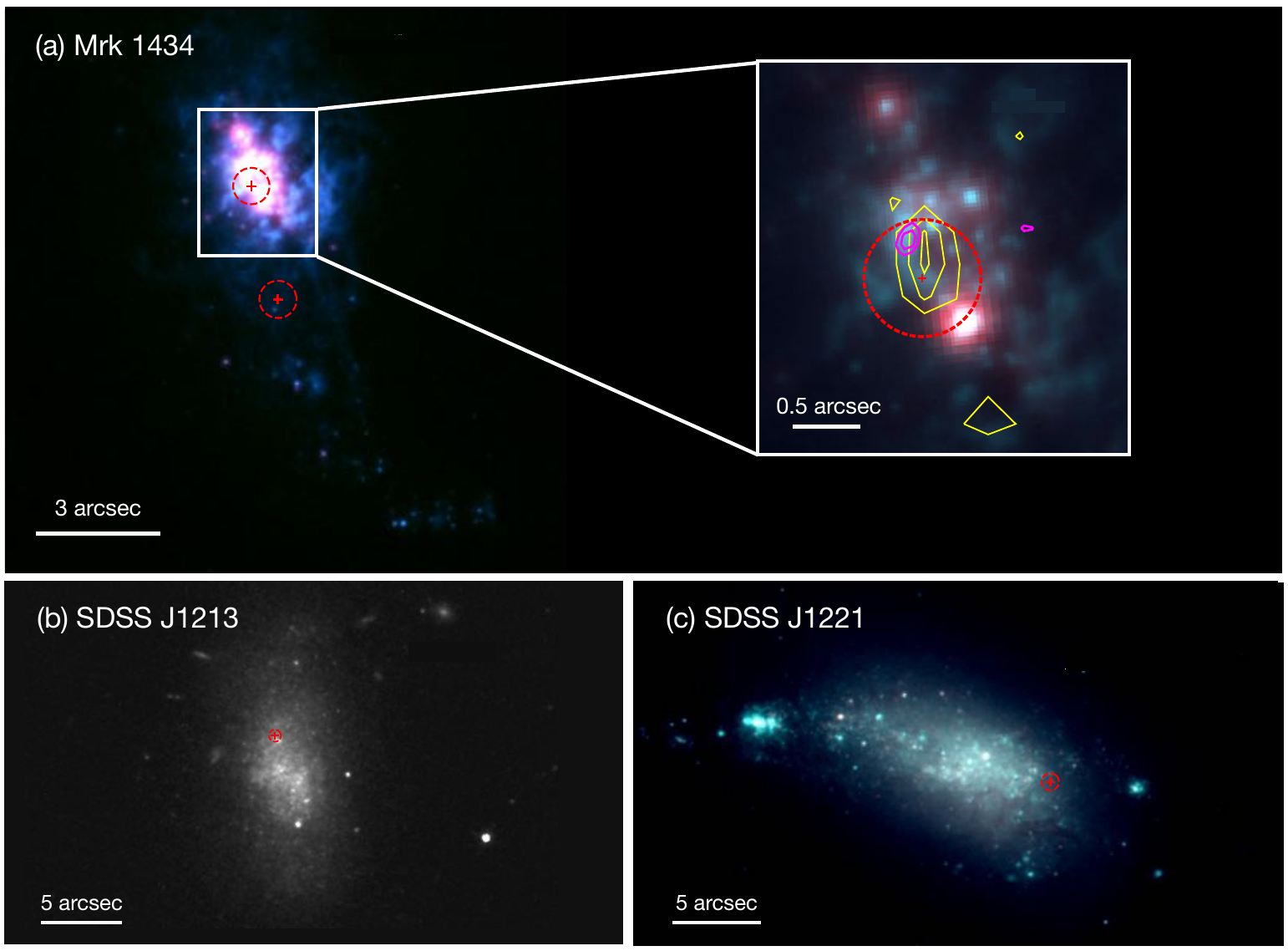}
    \caption{(a) Composite \textit{HST} image of \galone\ in the F606W (blue/green) and the F110W (red) filters. The locations of the two X-ray point sources are shown as red cross hairs, with the dashed red circles illustrating the sizes of the 95\% positional errors from \textit{Chandra}. The Zoom-in of the centre of the galaxy shows the location of \mrkn\ relative to the radio emission, where yellow contours show the extended 5.5 GHz radio emission ($1\farcs1\times0\farcs6$; contours drawn at 3, 4, 5$\times$$\sigma_{\rm rms}$) and the magenta contours show the unresolved emission at 9.0 GHz (contours drawn at 3, 4$\times$$\sigma_{\rm rms}$). The sizes of the VLA synthesised beams are $0\farcs45 \times 0\farcs38$ (5.5 GHz) and $0\farcs27 \times 0\farcs24$ (9.0 GHz), respectively. Note, the SDSS spectroscopic fibre, from which the nebular \ion{He}{ii} emission is detected, has a diameter of 3$\arcsec$ and is placed at the centre of the galaxy.  (b) \textit{HST} image of \galtwo\ in the F110W filter, with the location of the X-ray source marked by the red cross hair and dashed circle.  (c) \textit{HST} composite image of \galthree\ in the F606W (blue/green) and the F110W (red) filters, with the location of the X-ray source marked by the red cross hair and dashed circle. In all images, north is up and east is to the left.}
    \label{fig:galaxies}
\end{figure*}

In the following subsections we present the multiwavelength results for each galaxy, deferring discussions regarding the possible nature of each X-ray source to Section~\ref{sec:disc}.  Composite \textit{HST} images are shown for each galaxy in Figure~\ref{fig:galaxies}, including the locations of  X-ray sources.

\subsection{Mrk 1434} 
\label{sec:res:mrk1434}

Mrk 1434 hosts two X-ray sources separated by 2\farcs8 (see Figure~\ref{fig:galaxies}a), both of which are classified as ULXs: the northern source (\mrkn), which is located toward the galactic nucleus, has  an unabsorbed hard X-ray luminosity $L_{\rm 2-10\,keV}=\left(1.2\pm0.6\right)\times10^{40}$ erg s$^{-1}$, and the southern source (\mrks) has $L_{\rm 2-10\,keV}=\left(5.8\pm0.2\right)\times10^{39}$ erg s$^{-1}$.  The X-ray spectra of each source are fit by powerlaw models with photon indices of  $\Gamma=1.3\pm0.4$ for \mrkn\ and $\Gamma=1.7 \pm 0.4$ for \mrks.  Neither source shows evidence for significant intrinsic absorption. 

It is unlikely that either hard X-ray source is a  superposed foreground/background object. Given the density and flux distribution of hard X-ray sources in the cosmic X-ray background (see, e.g., Equation 2 of \citealt{moretti03}), we expect to only  find 0.001 and 0.003   hard X-ray sources with 2-10 keV fluxes similar (or brighter) than \mrkn\ and \mrks, respectively, within the projected size of the galaxy (which we conservatively approximate as a circle with a 20$\arcsec$ radius).  

Radio emission is detected only from the northern source,  \mrkn.  At 5.5 GHz, the emission is extended with major and minor axis full width half maxima of 1\farcs1$\times$0\farcs6 (160 pc $\times$ 90 pc), respectively, covering $\approx$3.5 synthesised beams.  The centroid of the 5.5 GHz emission is 0\farcs16  from the X-ray position (for reference, the 95\% \textit{Chandra} error circle is 0\farcs35), and the integrated luminosity is $L_{\rm 5.5,int}=\left(1.2\pm0.2\right)\times10^{36}$ erg s$^{-1}$.   At 9.0 GHz we detect a point source located 0\farcs32 from the X-ray position, with a peak luminosity $L_{\rm 9.0, peak} = \left(3.7\pm0.8\right)\times10^{35}$ erg s$^{-1}$.  We do not detect any extended radio structures at 9.0 GHz, thereby indicating that the emission seen at 5.5 GHz has a steep radio spectrum (our 5.5 and 9.0 GHz radio maps have similar sensitivities; see Table~\ref{tab:vla}).  Note,  extended emission is not simply resolved out at the higher radio frequency, since the smallest baselines of the VLA in  A configuration are sensitive to structures up to $\approx$5$\arcsec$ at 9.0 GHz, which is larger than the $\approx$1$\arcsec$ angular size of the 5.5 GHz emission. 

   At 9.0 GHz, the chance of a random alignment of a background radio point source falling within the \textit{Chandra} error circle is very small.  Integrating the differential source counts  tabulated by \citet{dezotti10} at 8.4 GHz, and assuming a flat radio spectrum (as expected if the 9.0 GHz  emission is from a compact jet; see Section~\ref{sec:disc:mrk1434:agn}), we expect only $\approx$$3\times10^{-5}$ sources with $f_{\rm peak}>0.036$ mJy within the X-ray error circle.   The chance of a statistical fluctuation as large as 0.036 mJy (i.e., 4$\sigma_{\rm rms}$) within the X-ray error circle (which contains $\approx$240 pixels in the radio map) is also very small ($p=3\times 10^{-5}$). Thus, we believe the 9.0 GHz emission is indeed physically associated with the  galaxy.  However, we note that the radio source lies toward the edge of the X-ray error circle.  Thus, even though the radio source formally falls within the \textit{Chandra} positional uncertainty,  its association specifically with \mrkn\ is less clear, particularly after considering that the \textit{Chandra} X-ray astrometry of \galone\ was aligned to the optical frame using only two common X-ray/SDSS sources.

Finally, we note that towards the southwest of the 0\farcs35 \textit{Chandra} X-ray error circle of \mrkn, there is an optical/near-infrared source that appears red in the \textit{HST} composite image (see the zoom-in of Figure~\ref{fig:galaxies}a).  If that source is a background quasar it may also be responsible for the X-ray and/or radio emission.  However, the random alignment of such a background quasar is very unlikely, as described below.   The  AB magnitude of the \textit{HST} source in the F606W filter is 18.8, which we convert to SDSS i$\approx$18.7 assuming a typical quasar spectrum \citep{vandenberk01}.  We then consider SDSS Type 1 quasar counts from $0.3 < z < 3.5$ \citep{richards06, ross13}, and we find only a negligible number of background quasars ($\approx6\times10^{-7}$) are likely to fall within the \textit{Chandra} X-ray circle by random chance (note, the random alignment of a radio-loud or a Type 2 quasar would be even rarer).  That source is likely intrinsic to the galaxy.

\subsection{\galtwo\ and \galthree}
\galtwo\ and \galthree\ each contain a single hard X-ray point source near the outskirts of each galaxy (Figure \ref{fig:galaxies}b-c). The hard (2-10 keV) X-ray luminosities of the sources are $L_{\rm 2-10\,keV}=\left(4.3\pm2.4\right)\times10^{39}$ and $\left(2.9\pm0.8\right)\times10^{39}$ erg s$^{-1}$, respectively (Table \ref{tab:chandra}), such that both sources are classified as ULXs.  The chance of a superposed foreground/background object is negligible (we expect only 0.005 hard X-ray background sources for \galtwo\ and 0.001 sources for \galthree; \citealt{moretti03}).  Neither galaxy contains radio emission within the \textit{Chandra} X-ray circles to 3$\sigma_{\rm rms}$ upper limits of 
$<1.4\times10^{35}$ erg s$^{-1}$ at 10.0 GHz for \galtwo, and to limits of 
$< 3.0 \times 10^{34}$ erg s$^{-1}$ and 
$<4.9\times10^{34}$ erg s$^{-1}$ at 5.5 and 9.0 GHz, respectively, for \galthree.  

\section{Discussion}
\label{sec:disc}

In the following subsections we discuss possible  interpretations for the X-ray sources in our sample of three dwarf galaxies. We focus primarily on \galone\ since it exhibits the most complex phenomenology (i.e., two X-ray sources, one of which is coincident with radio emission).  We provide arguments for/against XRB interpretations in Section \ref{sec:disc:mrk1434:xrb} and for/against AGN interpretations in Section \ref{sec:disc:mrk1434:agn}.  In Section \ref{sec:disc:mrk1434:heii} we discuss whether the observed X-ray flux is sufficient to explain  \ion{He}{ii} line emission observed in the SDSS spectrum of \galone.   A discussion on the nature of the X-ray sources in the other two galaxies is presented in Section \ref{sec:disc:othergals}.

\subsection{\galone}
\label{sec:disc:mrk1434}
\subsubsection{XRB Interpretations}
\label{sec:disc:mrk1434:xrb}

As shown in Section~\ref{sec:res:mrk1434}, both X-ray sources in \galone\ are physically associated with the galaxy and luminous enough to be classified as ULXs.  The observed X-ray luminosity, however, is higher than expected from  the luminous tail of the galaxy's XRB population. The luminosities of both X-ray sources are above the cutoff of the low-mass XRB luminosity function \citep[e.g.,][]{gilfanov04}, so in the following we only consider high-mass XRBs using the metallicity-dependent luminosity function from \citet{lehmer21}.   For \galone, with $12+\log\left(O/H\right)=7.8$ and SFR=0.12 $M_\odot$ yr$^{-1}$, \citet{lehmer21} predict a total 0.5-8.0 keV X-ray luminosity (i.e., from all X-ray point sources) of $L_{0.5-8.0\,{\rm keV}} = \left(1.7\pm0.15\right) \times 10^{39}$ erg s$^{-1}$ (where the error bar represents the 68\% confidence interval provided by \citealt{lehmer21}).  They also predict only $0.03^{+0.04}_{-0.02}$ ULXs with $L_{0.5-8.0} > 10^{40}$ erg s$^{-1}$.   For reference, the unabsorbed 0.5--8.0 keV model luminosities of \mrkn\ and \mrks\ are $\left(1.3\pm0.4\right)\times 10^{40}$ and $\left(0.8\pm0.2\right)\times10^{40}$ erg s$^{-1}$, respectively.  Thus, the combined X-ray luminosity of both ULXs is $\approx$10 times higher than expected relative to the \citet{lehmer21} luminosity function, which is significant even after considering uncertainties and intrinsic scatter. 

Even though the above suggests that it is statistically unlikely for both sources to be XRBs, small number statistics could influence the above arguments, and it is  worth  exploring XRB interpretations.  In particular, the extended 5.5 GHz radio emission from \mrkn\ could represent a `ULX bubble', as similar types of extended radio structures have been observed from other ULXs,  making the radio emission a signature of a ULX outflow shocking the nearby interstellar environment  \citep[e.g.,][]{pakull10, soria10, soria21, cseh12, urquhart19}.  If the 5.5 GHz radio emission is indeed a ULX bubble, then with $L_{\rm 5.5,int}=\left(1.2\pm0.2\right)\times10^{36}$ erg s$^{-1}$ it would represent the most luminous ULX bubble yet observed by a factor of $\approx$6 \citep{pakull10, soria10, soria21}. Meanwhile, the projected size of $\approx$160 pc $\times$ 90 pc ($1\farcs1 \times 0\farcs6$) in diameter is fairly typical compared to other ULX bubbles, where diameters range from $\approx$25--350 pc (\citealt{soria21}; also see Table~1 of \citealt{berghea20} and references therein).  Taking the peak flux density of the 5.5 GHz structure, and extrapolating to 1 GHz assuming a spectral index $\alpha=0.7$, the intensity of the  radio bubble in \mrkn\ would be $I_{\rm 1\,GHz}\approx 6\times10^{-16}$ erg s$^{-1}$ cm$^{-2}$ Hz$^{-1}$ sr$^{-1}$, which is relatively large but reasonable compared to other ULX radio bubbles with similar physical sizes (see Figure~5 of \citealt{berghea20}).

Although a ULX bubble is one interpretation of the 5.5 GHz emission, we stress that it is not a unique (or necessary) explanation. Adopting SFR $=0.12$ $M_\odot$ yr$^{-1}$ for \galone\ and the relation between star formation rate and the 1.4 GHz specific luminosity from \citet{kennicutt12}, we expect $L_{\rm 5.5,SF} \approx 3.9\times10^{36}$ erg s$^{-1}$  (we convert from 1.4 GHz to 5.5 GHz assuming a spectral index $\alpha=0.7$).  Considering that the intrinsic scatter on the conversion between SFR and radio luminosity is on the order of $\pm$0.3 dex \citep{murphy11}, the observed extended structure at 5.5 GHz could be produced entirely by star formation processes.   Since the extended radio structure at 5.5 GHz is not detected at 9.0 GHz, the dominant radio emission mechanism in such a scenario would most likely  be synchrotron radiation with a steep spectrum from supernova remnants.  Note, our data exclude free-free radio emission from an \ion{H}{ii} region, which would produce a flat spectrum that would be detectable at 9.0 GHz. 

\subsubsection{AGN Interpretations}
\label{sec:disc:mrk1434:agn}

AGN can also produce extended radio emission, which is another viable explanation for the 5.5 GHz radio structure.  However, in light of the discussion in the previous subsection that a super-Eddington XRB is also capable of producing the observed extended emission at 5.5 GHz, the resolved radio complex does not provide useful diagnostics for attempting to discriminate between XRB vs.\  AGN.  Since the X-ray spectra of \mrkn\ and \mrks\ ($\Gamma=1.3\pm0.4$ and $\Gamma=1.7 \pm 0.4$, respectively) are consistent with low-luminosity AGNs \citep{younes11, yang15}, we focus the following discussion on AGN scenarios with Eddington ratios $L_{\rm bol}/L_{\rm Edd} \lesssim 0.01$.  For such weakly accreting AGN,  we expect  to observe unresolved radio emission from a partially self-absorbed compact jet \citep{ho08}. By combining X-ray and radio luminosities,  we can then make crude estimates on  black hole masses by appealing to the fundamental plane of black hole activity \citep{merloni03, falcke04}.  For \mrkn, we then interpret the the unresolved 9.0 GHz radio emission as arising from a compact jet, and we
utilise the fundamental plane regression by \citet{gultekin19},
\begin{equation}
\begin{aligned}
    \log\left(M_{\rm BH}/10^8 M_\odot\right) = & \left(0.55\pm0.22\right) + \\ & \left(1.09\pm0.10\right)\log\left(L_{\rm 5\,GHz}/10^{38}\,{\rm erg\,s}^{-1}\right) - \\ &
\left(0.59\pm0.16\right)\log\left(L_{\rm 2-10\, keV}/10^{40}\,{\rm erg\,s}^{-1}\right),
\end{aligned}
\end{equation}
which has an intrinsic scatter $\approx$1 dex.  We estimate that \mrkn\ would have $M_{\rm BH} \approx 4 \times 10^{5}~M_\odot$ if powered by an mBH (see Table~\ref{tab:masses}).  Note, we assume a flat radio spectrum to convert the observed radio luminosity at 9.0 GHz to 5.0 GHz for use in the fundamental plane (we cannot use our 5.5 GHz radio map to estimate the 5 GHz luminosity because we do not have enough signal-to-noise to attempt to decompose a point source embedded within the extended radio emission observed at 5.5 GHz).  Similarly, the lack of radio emission from \mrks\ implies $M_{\rm BH} \lesssim 4 \times 10^5 M_\odot$ (where we adopt a 3$\sigma_{\rm rms}$ upper limit, based on the observed $\sigma_{\rm rms}$ near \mrks\ in our 5.5 GHz image).  These mass estimates imply Eddington ratios ($L_{\rm 2-10\,keV}/L_{\rm Edd}$) of $\approx 2 \times10^{-4}$ and $\gtrsim 1\times10^{-4}$ for \mrkn\ and \mrks, respectively, which, assuming bolometric corrections of $\approx$10, are consistent with Eddington ratios for which the fundamental plane can be applied \citep[see, e.g.,][]{plotkin12}.

\begin{table}
    \centering
    \caption{mBH mass estimates and limits. %
    Column 1: galaxy name. %
    Column 2: logarithm of the hard X-ray luminosity. %
    Column 3: logarithm of the radio luminosity at 5 GHz, assuming a flat radio     spectrum.  For \mrkn, this luminosity is based on the unresolved emission     detected at 9 GHz.  For all other X-ray sources, limits are placed as     3$\sigma_{\rm rms}$. %
    Column 4: logarithm of the black hole mass (or limit) if X-ray sources are     weakly accreting mBHs, based on the fundamental plane of black hole activity     \citep{gultekin19}.  Uncertainties on $\log M_{\rm BH}$ are $\approx$1 dex.}
    \label{tab:masses}
    \renewcommand{\arraystretch}{1.} 
    \begin{tabular}{cccc}
        \hline
        Source & $\log L_{\rm 2-10\,keV}$  & $\log L_{\rm 5\,GHz}$ & $\log M_{\rm BH}$ \\  
               &  (erg s$^{-1}$)           &   (erg s$^{-1}$)     & ($M_\odot$)     \\ 
        (1)  &  (2)  & (3)           &  (4) \\
         \hline
\mrkn & $40.1\pm0.4$ & $35.3\pm0.1$ & 5.6 \\
\mrks & $39.8\pm0.3$ & $<$35.2 & $<$5.6 \\
\galtwo & $39.6\pm0.4$ & $<$34.9 & $<$5.3\\
\galthree & $39.5\pm0.2$ & $<$34.4 & $<$5.0 \\
         \hline
    \end{tabular}
\end{table}

\subsubsection{On the Origin of Nebular \ion{He}{ii} Emission}
\label{sec:disc:mrk1434:heii}

In the following we determine whether the X-ray emission from \galone\ is a strong enough source of photoionisation to explain the  strength of the \ion{He}{ii} emission in the SDSS spectrum of \galone. The observed \ion{He}{ii} line flux is $F_{\rm 4686, obs} = \left(7.5\pm0.1\right)\times10^{-16}$ erg s$^{-1}$ cm$^{-2}$, which translates to a photon flux of $N_{\rm 4686, obs} = \left(1.8\pm0.1\right)\times10^{-4}$ photons s$^{-1}$ cm$^{-2}$. Every photon emitted in the \ion{He}{ii} line requires 5.2 ionizing photons incident on singly ionised helium \citep{pakull86}.  Given the ionisation potential of singly ionised helium ($\chi_{\rm ion}=54.4$ eV), and considering that the photoionisation cross section has a steep $E_{\rm ph}^{-3}$ dependence on photon energy, $E_{\rm ph}$, then producing the observed SDSS \ion{He}{II} line flux requires a photon flux in the extreme ultraviolet  (54--300 eV) of $N_{\rm 54-300\,eV} = 5.2 N_{\rm 4686, obs} = \left(9.1\pm0.1\right)\times10^{-4}$ photons s$^{-1}$ cm$^{-2}$.  Note, this photon flux is an underestimate because we have not corrected the observed SDSS line flux for extinction.

The 3$\arcsec$ SDSS spectroscopic fibre is centred near \mrkn, such that if the \ion{He}{II} emission arises from photoionisation by the X-ray source, we expect the emission to be dominated by \mrkn.  We do not have  direct measurements on the extreme ultraviolet flux from 54-300 eV, so we  extrapolate the \textit{Chandra} X-ray spectrum into the extreme ultraviolet.  Our best-fit powerlaw model predicts a photon flux of $0.3^{+2.5}_{-0.2}\times10^{-4}$ photons s$^{-1}$ cm$^{-2}$ (note the large range in uncertainty because we are extrapolating the model to energies lower than the \textit{Chandra} X-ray band).  Thus, while high-energy radiation from \mrkn\ may contribute to some of the \ion{He}{II} photoionisation, the observed X-ray source is too faint, by a factor of $\approx$30, to supply all of the photoionising photons. If we assume a thermal X-ray emission model ({\tt tbabs*diskbb}), it becomes even more difficult for the X-ray source to explain the \ion{He}{II} photionisation, as the extrapolated 54-300 eV extreme ultraviolet flux becomes $\approx$90 times too faint. Adding a contribution of photons form \mrks\ would only increase the above photon flux by a factor of $\approx$2, for either spectral model.

There is currently no evidence for significant X-ray variability from \galone\ over the past 1--2 decades. Coincidentally, the SDSS spectrum and the archival \textit{Chandra} observation from \citet[][\textit{Chandra} obsID 3347]{lemons15} were both taken in May 2002 (separated by $\approx$2 weeks).   The archival data from 2002 show nearly identical  X-ray luminosities ($\log L_{2-10\,{\rm keV}} = 40.1$ and 39.9 erg s$^{-1}$ for \mrkn\ and \mrks, respectively; see Table~2 of \citealt{lemons15}) compared to the  \textit{Chandra} observations presented here, which were taken nearly 14 years later (see Table~\ref{tab:chandra} of this paper).  There are also two X-ray detections of \galone\ in the third \textit{XMM-Newton} serendipitous source catalog \citep[3XMM;][]{rosen16} in 2007 and 2008.  Both X-ray sources are blended together due to \textit{XMM-Newton}'s poorer spatial resolution.  Comparing the \textit{XMM-Newton} fluxes to the combined fluxes of both sources in the \textit{Chandra} observations, X-ray variability is smaller than a factor of $\approx$2 over the four observations. However, considering the light travel time between the X-ray source and the ionised medium, it is feasible that \mrkn\ was more active  in the past.  The projected radius of the SDSS spectroscopic fibre is 730 light years, and we cannot exclude the possibility that \mrkn\ was $\approx$30--90 times more luminous several hundred years ago, which appears to be on the only viable way for the \ion{He}{II} emission to be powered by X-ray photoionisation.

If the extended radio emission is  produced by an outflow shocking the interstellar medium, then one must also consider the possibility of the \ion{He}{II} emission being produced by ionisation  from a radiative shock \citep[e.g.,][]{dopita95}. According to the {\tt MAPPINGS III} libraries of line ratios for radiative shocks \citep{allen08}, assuming a shock velocity of 300 km s$^{-1}$, we expect the luminosity of the \ion{He}{ii} $\lambda$4686 emission line $L_{4686} \approx 4\times10^{-4} L_{\rm rad}$, where $L_{\rm rad}$ is the total radiative luminosity of the shock.\footnote{Given the low metallicity of \galone, we adopt the {\tt MAPPINGS III} model grid with Small Magellanic Cloud abundances.  We also assume an interstellar medium density of 1 cm$^{-3}$ and equipartition of  magnetic and thermal pressures.}  
  Assuming that the kinetic power required to inflate a bubble $P_{\rm kin} \approx 77/27 L_{\rm rad}$ \citep{weaver77}, then explaining the observed \ion{He}{II} line via shock ionisation requires an outflow with $P_{\rm kin} \approx 6 \times 10^{41}$ erg s$^{-1}$.

 We do not have a reliable method to independently estimate $P_{\rm kin}$ (especially considering that other emission lines in the SDSS spectrum are  dominated by star formation).  However, for an order of magnitude estimate, we calculate the minimum synchrotron energy of the 5.5 GHz radio emission, which is $W_{\rm min} \approx 2\times10^{52}$ erg \citep{longair94}.\footnote{We adopt $L_{\rm 5.5} \approx 10^{36}$ erg s$^{-1}$, a bubble diameter of $\approx$160 pc, and an ion to electron energy ratio of $\eta=40$.  We note that  $W_{\rm min} \propto \eta^{4/7}$, and the proper value of $\eta$ is not well constrained.} 
 A 300 km s$^{-1}$ shock would take $\approx 3\times10^5$ yr to inflate a bubble with a 160 pc diameter, such that the average  power stored in internal energies of the synchrotron emitting structure  is $\bar{P}_{\rm min} \approx 2\times 10^{39}$ erg s$^{-1}$ (i.e., the average power in particles and in the magnetic field). Thus, an outflow would  need to carry $\gtrsim$10$^2$ times more power in order for a shock to be the sole ionisation source of the observed \ion{He}{II} emission line.  Of course, $\bar{P}_{\rm min}$ is a minimum energy estimate, and the power in bubbles/cavities carved out by kinetic outflows have sometimes been observed to be larger, sometimes by factors of several hundreds \citep[e.g.,][]{ito08}, such that the above does not exclude the possibility of shock ionisation. 

 For comparison, the ULX NGC 6946 MF16 \citep{roberts03} has a luminous and compact radio bubble \citep{berghea20}, which suggests a relatively powerful outflow. Adopting the NGC 6946 MF16 bubble line flux in the [\ion{Fe}{II}] $\lambda$16440 emission line ($4.2\times10^{-15}$ erg s$^{-1}$ cm$^{-2}$) and a distance of 7.8 Mpc \citep{long20}, the {\tt MAPPINGS III} libraries for a 300 km s$^{-1}$ shock (with Solar abundances) suggest a kinetic power of $P_{\rm kin}\approx 7 \times 10^{40}$ erg s$^{-1}$.  Thus, the kinetic power of NGC 6946 MF16 (i.e., one of the most powerful known ULX radio bubbles) is an order of magnitude lower than the power required for shock ionisation to be responsible for the observed strength of the \ion{He}{II} emission line near \mrkn.  Thus, if the \ion{He}{II} line is powered by shock ionisation, then it would represent one of the most powerful bubbles carved by a ULX outflow yet observed.

Intriguingly, \galone\ is one member of a population of 182 star forming galaxies with nebular \ion{He}{ii} emission that were identified by \citet{shirazi12}. The ratios of \ion{He}{ii}/H$\beta$ relative to [\ion{N}{ii}] $\lambda$6584/H$\alpha$ are inconsistent with AGN for these galaxies.     Typically, when an AGN is absent, Wolf-Rayet stars are considered the primary stellar population capable of producing enough extreme ultraviolet flux above the 54 eV \ion{He}{ii} ionisation edge.  However, \citet{shirazi12} inspected the SDSS spectra for broad emission features indicative of Wolf-Rayet stars, and they found no Wolf-Rayet signatures in the spectrum of \galone.  Thus, without concrete evidence that \mrkn\ was indeed brighter several hundred years ago to power the \ion{He}{II} emission via photoionisation, and/or lacking a reliable estimate of the kinetic power of an outflow for shock ionisation, the source of extreme ultraviolet photons in \galone\ remains a mystery.  Another plausible explanation could be photoionisation from extreme ultraviolet photons emitted by exotic stellar populations (like rapidly rotating stars) in metal-poor environments (see the discussion in Section 6 of \citealt{shirazi12}). It is very plausible that several of the above scenarios contribute toward producing the \ion{He}{II} line, and \citet{shirazi12} recovered a heterogeneous population (multiple mechanisms may even contribute to producing the \ion{He}{II} emission within a single galaxy).  For example,  \citet{senchyna20} conclude that X-ray photoionisation cannot explain nebular \ion{He}{II} emission across a sample of nearly a dozen  metal-poor galaxies.  Meanwhile, there are several well-established examples of X-ray sources that are indeed sufficient to power nebular \ion{He}{II} emission \citep[e.g.,][]{pakull86, moon11, schaerer19, simmonds21}.  Further observational constraints, ideally via systematic X-ray surveys of metal-poor dwarf galaxies under high spatial resolution,  are required to understand the level to which ULXs contribute extreme ultraviolet radiation in metal-poor galaxies, which has  implications for understanding sources of ionisation and heating of the intergalactic medium in the early Universe.

\subsection{\galtwo\ and \galthree}
\label{sec:disc:othergals}

Our new \textit{Chandra} observations confirm the conclusion of \citet{lemons15} that both X-ray sources are more luminous than expected from the XRB populations in each galaxy, as described below.    Unlike for \galone, the luminosities of both X-ray sources in \galtwo\ and \galthree\ are low enough that we should consider both high-mass and low-mass XRBs.  Following \citet{lemons15}, we therefore adopt the relation from \citet{lehmer10}, which predicts the hard X-ray luminosity from low-mass and high-mass XRBs as a function of stellar mass and star formation rate: $\left(L_{2-10}^{\rm XRB}/{\rm erg\,s^{-1}}\right) = \left(9.05\pm0.37\right)\times10^{28} \left(M_{\star}/M_\odot\right) + \left(1.62 \pm 0.22\right)\times10^{39}\left(SFR/{M_\odot\,{\rm yr}^{-1}}\right)$, with an intrinsic scatter of $\pm$0.34 dex.  The \citet{lehmer10} relation  predicts $L_{2-10}^{\rm XRB}=1.2\times10^{37}$ and $5.6\times10^{37}$ erg s$^{-1}$ for \galtwo\ and \galthree, respectively.  The predicted luminosities are $\approx$3 times higher if we instead adopt the calibrations in \citet{lehmer19}. Thus, the observed X-ray luminosities are $\approx$120--360 and $\approx$17--50 times higher than expected, for \galtwo\ and \galthree, respectively.\footnote{The \citet{lehmer10} relation is calibrated to galaxies with approximately Solar metallicities.  The metallicity of \galtwo\ is unknown, and the metallicity of \galthree\ is $\log\left(O/H\right)+12=8.3$ \citep{zhao13}.  If we adopt the metallicity-dependent \citet{lehmer21} relation for high-mass XRBs, the X-ray luminosity of the X-ray source in \galthree\ is still $\approx$20 times higher than expected for a galaxy with its star formation rate and metallicity.}  

In light of recent theoretical motivation for `wandering' mBHs \citep[][also see, e.g., \citealt{mezcua20, reines20, greene21, sargent22} for observational searches]{bellovary19, bellovary21}, an X-ray source being `off-nucleus' does not on its own preclude the possibility of an accreting mBH.  It is  possible that these sources are mBHs launching jets that are either (a) beneath our radio detection limit or (b) that are very extended and `resolved out' by the VLA when it is in its most extended A configuration.  The largest angular scale to which the VLA is sensitive to radio emission at our observing frequencies (X-band) and configuration (A) is $5\farcs3$, such that our VLA observations would not detect  flux from extended jets larger than $\approx$850 and $\approx$410 pc for \galtwo\ and \galthree, respectively.  On the other hand, the radio cores of  weakly accreting AGN (bolometric luminosities $L_{\rm bol}<0.01 L_{\rm Edd}$) have flat radio spectra and are compact enough that their radio emission should not be `resolved out' at VLA resolutions \citep[see, e.g.,][]{orienti10}.  Thus, if only considering mBHs in the weak accretion regime, we can use our radio upper limits in conjunction with the fundamental plane to place mass limits of  $M_{\rm BH}< 2\times10^{5}$ and $<$$1\times10^{5}~M_\odot$ for \galtwo\ and \galthree, respectively.  Requiring $L_{\rm bol}<0.01 L_{\rm Edd}$, and assuming  X-ray bolometric corrections of 10, then places lower limits on black hole masses of $\gtrsim3\times10^4$ (\galtwo) and $\gtrsim2\times10^4\,M_\odot$ (\galthree). Thus, there is a relatively narrow range of mass where our VLA observations could `miss' the compact radio jet from a   weakly accreting mBH.  Note, our radio limits do not place useful constraints on the possibility of a more rapidly accreting mBH with $L_{\rm bol}>0.01 L_{\rm Edd}$, which would correspond to a mass $M_{\rm BH} \lesssim 10^4 M_\odot$ for both sources.  Nevertheless, even though our data do not exclude the possibility of  mBHs,  Occam's razor probably suggests that the simplest and most likely scenario is that these are luminous XRBs.

\subsection{An Update to Lemons et al.\ (2015)}
\label{sec:disc:lemons}

After considering the above multiwavelength observations, all 10 of the dwarf galaxy AGN candidates identified by \citet{lemons15} (via hard X-ray emission) now have sufficient spatial resolution to determine if the X-ray sources indeed reside in  galactic nuclei.  Our study reduces their number of AGN candidates to 7--8 (adopting an AGN definition that requires nuclear sources). It is very unlikely that any of these 7--8 nuclear sources are chance alignments with foreground/background X-ray emitting objects.  Adopting the hard (2-10 keV) X-ray fluxes and X-ray position error circles of the nuclear candidates from Table 2 of \citet{lemons15}, and replacing the X-ray flux and positional uncertainty of \mrkn\ with the values presented here, the \citet{moretti03} cosmic X-ray background predicts only 0.003 sources to fall within the nuclei of the eight possible nuclear mBH candidates. Obtaining 7--8 viable AGN candidates is a significant result, considering that (a) the \citet{lemons15} dwarf galaxy survey was archival and therefore serendipitous in nature, and (b) the three dwarf galaxies with follow-up presented here represent three of their most unlikely AGN candidates (given the poor spatial resolution of their archival \textit{Chandra} data).  \citet{lemons15} found X-ray sources in 19 galaxies total (i.e., the remaining 11--12 galaxies host off-nuclear X-ray sources, most likely XRBs).  Thus, \textit{if} a luminous X-ray source is detected in a dwarf galaxy, our study (very roughly) implies a  30--40\% chance\footnote{This number is an upper limit, and it neglects biases inherent to an archival/serendipitous survey, which is out of the scope of this paper to quantify.}  
that it could be a nuclear mBH, which supports the viability of using X-ray surveys to identify mBHs in low-mass galaxies, as long as the survey is performed with sufficient sensitivity and spatial resolution.  We stress the importance of high spatial-resolution X-ray observations.  For example, \galone\ was previously  identified as an AGN from an \textit{XMM-Newton} survey \citep{birchall20}, while our higher spatial-resolution \textit{Chandra} observation clearly resolves the `nuclear' X-ray source into two distinct sources (and even then, it remains unclear if either source is indeed an accreting mBH).

\section{Summary and Conclusions} \label{sec:conc}

We have presented a multiwavelength study of three nearby dwarf galaxies that host ULXs.  Two galaxies in our sample, \galtwo\ and \galthree, each contain single off-nuclear X-ray sources that we suspect are luminous XRBs.  The third galaxy, \galone\ hosts two X-ray sources separated by 2\farcs8.  The northern source (\mrkn) also displays extended radio emission at 5.5 GHz and point-like radio emission at 9.0 GHz.   It remains unclear if the X-ray sources in \galone\ are XRBs or AGNs (especially \mrkn),  although either scenario is intriguing.  If XRBs, then the combined X-ray luminosity of both sources is larger than expected for a galaxy with \galone's star formation rate and (low) metallicity.  Futhermore, the extended radio emission at 5.5 GHz could then represent the most luminous `ULX bubble' ever observed in the radio,  although we stress that the 5.5 GHz radio emission can also be attributed entirely to star formation within the galaxy, or to an AGN jet. Regardless of the correct scenario, we find that the line emission from \ion{He}{ii} in \galone\ is inconsistent with a nebula being powered by the central X-ray source, unless the central source underwent a period of higher activity several hundred years ago, or if the the nebula is shock ionised by an outflow that is an order of magnitude more powerful than yet observed from a ULX. If \mrkn\ is an AGN, then the 9.0 GHz radio emission may represent a compact synchrotron jet from a low-luminosity AGN power by an mBH with $M_{\rm BH}\approx 4\times 10^5 M_\odot$. We conclude by stressing the importance of high spatial-resolution observations when performing  multiwavelength searches for mBHs in dwarf galaxies.


\section*{Acknowledgements}

We thank the anonymous referee for helpful comments that improved this manuscript. Support for this work was provided by the National Aeronautics and Space Administration through Chandra Award Number GO6-17079X issued by the Chandra X-ray Center, which is operated by the Smithsonian Astrophysical Observatory for and on behalf of the National Aeronautics Space Administration under contract NAS8-03060.  This research is based on observations made with the NASA/ESA Hubble Space Telescope obtained from the Space Telescope Science Institute, which is operated by the Association of Universities for Research in Astronomy, Inc., under NASA contract NAS 5–26555. These observations are associated with program HST-GO-14356. Support for Program No. HST-GO-14356 was provided by NASA through a grant from the Space Telescope Science Institute, which is operated by the Association of Universities for Research in Astronomy, Incorporated, under NASA contract NAS5-26555.  RMP and JDP acknowledge support from the National Science Foundation under grant No.\ 2206123.  RS acknowledges support from grant number 12073029 from the National Natural Science Foundation of China (NSFC). AER acknowledges support provided by NASA through EPSCoR grant number 80NSSC20M0231. GEA is the recipient of an Australian Research Council Discovery Early Career Researcher Award (project number DE180100346) funded by the Australian Government.  This research made use of Astropy,\footnote{\url{http://www.astropy.org}} a community-developed core Python package for Astronomy \citep{astropy13, astropy18}.

\section*{Data Availability}

 The data underlying this article are available in the Chandra Data Archve under ObsIDs 18059, 18060, and 18061 (\url{https://cda.harvard.edu/chaser/}), in the Barbara A. Mikulski Archive for Space Telescopes  under program ID 14356 (\url{dx.doi.org/10.17909/3bxp-zt07}), and in the National Radio Astronomy Observatory Data Archive under programs 14-358 and SH0563 (\url{data.nrao.edu}).



\bibliographystyle{mnras}
\bibliography{references} 







\bsp	
\label{lastpage}
\end{document}